# Phenomenological renormalizaiton group approach to the anisotropic two-layer Ising model

B. Mirza[†‡] and T. Mardani[†]

[†] *Department of Physics, Isfahan University of Technology, Isfahan 84154, Iran*

[‡] *Institute for Studies in Theoretical Physics and Mathematics,*

*P.O.Box 5746, Tehran, 19395, Iran*

*E-mail: b.mirza@cc.iut .ac.ir*

**ABSTRACT**

The anisotropic two-layer Ising model is studied by the phenomenological renormalizaiton group method. It is found that the anisotropic two-layer Ising model with symmetric couplings belongs to the same universality class as the two dimensional Ising model.

I. **Introduction**

For many years, the lattice statistics has been the subject of intense research interests. Although, at zero magnetic field, there is an exact solution for the 2-dimensional Ising model [1], however, there is no such a solution for the two-layer Ising model. The two-layer Ising model as a simple generalization of the 2-D Ising model has been studied for a long time [2-6]. The two-layer Ising model as a simple model for the magnetic ultra-thin film has various possible applications to real physical materials. It has been found that capping PtCo in TbFeCo to form a two-layer structure has applicable features, for instance, raising the Curie temperature and reducing the switching fields for magneto-optical disks [7]. Cobalt films grown on a Cu (100) crystal have highly anisotropic magnetization [8] and could be viewed as layered Ising models. In recent years, some approximation methods have been applied to this model [9-17]. A critical line has been found in all these studies. As is expected the Curie temperature is very sensitive to the inter-layer interaction. Many discussions have been directed to the shift exponent at the decoupling point. Abe [3] and Suzuki [4] have predicted $\gamma = 7/4$ for the isotropic model many years ago. Recent computational results are in agreement with earlier results [14,15,18]. Apart from the shift exponent, it is also interesting to study the critical behaviour along the critical line. The model has the same critical exponent at the two ends of the critical line corresponding to the solvable decoupling limit and the infinite interlayer coupling limit. But it is clear that the decoupled system has a higher symmetry than the coupling layers, hence one cannot assume a priori that the two layer Ising model belongs to the one-layer Ising universality class. It has been proposed that the critical exponents would vary continuously along the critical line [15]. However, there are also arguments in favour of unchanged exponents [11,26]. The question of the universality class of the two coupled, identical Ising layers was essentially settled in the seventies by van Leeuwen [26] who presented a scaling argument to explain why the Baxter 8-vertex model has continuously varying exponents. The same argument and the symmetry of the order parameter suggest that a system consisting of two coupled Ising model is likely to be in the universality class of the two-dimensional Ising model, however, the scaling arguments obviously are not rigorous and a numerical verification is useful. It is our purpose to provide a reliable prediction for the critical exponents based on the phenomenological renormalizaiton group method [19]. The term 'phenomenological renormalizaiton group' is sometimes used to denote a technique due to Nightingale [20] that is particularly powerful in the case of two-dimensional systems for which one can construct a transfer matrix. This technique has been applied with great success to the eight-vertex model [21], the square Ising antiferromagnet [22], the hard-square lattice gas [23], and the triangular Ising antiferromagnet [24]. In this method, it is essential to compute the largest and the next largest eigenvalues of transfer matrix.

## II. **Method**

For the two-layer Ising model with nearest neighbor interactions we are able to construct the transfer matrices and use it to calculate the critical exponents. Consider a two-layer square lattice with the periodic boundary condition composed of slices, each with two layers, each layer with $p$ rows, where each row has $r$ sites. Each slices has then $2 \times p \times r = N$ sites and the coordination number of all sites is the same (namely, 5). In the two- layer Ising model, for any site we define a spin variable

$$\sigma^{1(2)}(i+r, j) = \sigma^{1(2)}(i, j) \qquad (1)$$

$$\sigma^{1(2)}(i, j+p) = \sigma^{1(2)}(i, j) \qquad (2)$$

In this paper, we discuss the anisotropic ferromagnetic case in a magnetic field with the nearest neighbor coupling $J_x$ and $J_y$, where $J_x$ and $J_y$ are the nearest neighbor interactions within each layer in the x and y directions, respectively, and with inter-layer coupling $J_z$. We take only the interactions among the nearest neighbors into account. The configuration energy for the model may be defined as,

$$\frac{E(\sigma)}{kT} = -\sum_{i=1}^{r}\sum_{j=1}^{p}(\sum_{n=1}^{2}[J_x\sigma^n(i,j)\sigma^n(i+1,j) + J_y\sigma^n(i,j)\sigma^n(i,j+1)] \\ + J_z\sigma^1(i,j)\sigma^2(i,j) + h(\sigma^1(i,j) + \sigma^2(i,j))) \quad (3)$$

The canonical partition function, $Z(J)$, is

$$Z(J) = \sum_{\{\sigma\}} e^{\frac{-E(\sigma)}{kT}} \quad (4)$$

The transfer matrix element connecting the stripes $s_1, s_2, s_3, s'_1, s'_2, s'_3$ and $t_1, t_2, t_3, t'_1, t'_2, t'_3$ then can be written in the form

$$\langle s_1, s_2, s_3, s'_1, s'_2, s'_3 | T | t_1, t_2, t_3, t'_1, t'_2, t'_3 \rangle = \exp[J_x(s_1s_2 + s_2s_3 + s_3s_1 + t_1t_2 + t_2t_3 + t_3t_1) \\ + J_y\sum_{i=1}^{3}(s_is'_i + t_it'_i) + J_Z\sum_{i=1}^{3}(s_it_i) + h\sum_{i=1}^{3}(s_i + t_i)] \quad (5)$$

where $J_x, J_y$ and $J_z$ are the couplings and $h$ is a magnetic field.

The transfer matrix is a $2^{2n} \times 2^{2n}$ matrix which for the above example it is a $64\times64$ matrix and its largest eigenvalue can be used to obtain the partition function of a semi-infinite strip $(2\times n\times\infty)$. In particular the correlation length $\xi_{2n}$ is given by

$$\xi_m^{-1}(J,h) = \text{Log}\left|\frac{\Lambda_0^{(m)}}{\Lambda_1^{(m)}}\right| \quad , \quad m = 2n \quad (6)$$

where $\Lambda_0^{(m)}$, $\Lambda_1^{(m)}$ are the largest and the next largest eigenvalues of T in magnitude. Over two decade ago [19] the thermodynamic function of such semi-infinite stripes (in two dimensions) were calculated from the largest eigenvalues of $T$. Although the thermodynamic functions of these strips were informative, it was difficult to extract from them the critical exponents of an infinite system. Nightingale [20] showed how this could be achieved by viewing the correlation functions $\xi_m(J,h)$ and $\xi_{m'}(J',h')$ of two strips of different widths as being related by renormalization group transformation. The scaling of such a transformation implies

$$\frac{\xi_m(J,h)}{m} = \frac{\xi_{m'}(J',h')}{m'} \tag{7}$$

Strictly speaking, scaling implies an equation like this with the same function $\xi_m(J,h)$ on both sides. Eq. (7) with different functions $\xi_m$ and $\xi_{m'}$ on both sides is to be viewed as an approximate renormalization-group transformation, one which improves as the width of the strips increase. Both correlation functions $\xi_m$ and $\xi_{m'}$ can be calculated from Eq.(6) so that Eq.(7) and one additional assumption define the recursion relations

$$J' = J'(J,h) \quad , \quad H' = H'(J,h) \tag{8}$$

Which can be solved for fixed point values and linearized about them to obtain the critical exponents in the standard way. The problem now reduces to calculating the eigenvalues of $T_{kl}$ for a semi-infinite strip of width $m = 2n$. The correlation function invariably occurs in what follows in the combination $\xi_m/m$ so it is convenient to denote this by a single symbol

$$\zeta_m(J,h) = \xi_m(J,h)/m \tag{9}$$

We first consider the basic relation of Eq. (7) here written

$$\zeta_{m'}(J',h') = \zeta_m(J,h) \qquad (10)$$

in zero magnetic field. The correlation length for two different strip widths is calculated and a fixed point $J_C$ determined from

$$\zeta_{m'}(J',0) = \zeta_m(J,0) \qquad (11)$$

By linearizing Eq. (10) in $J$ about $J_C$ for $h = 0$ we obtain the thermal exponent $Y_t$ from

$$\left[\frac{m}{m'}\right]^{Y_t} = \frac{\partial \zeta_m/\partial J}{\partial \zeta_{m'}/\partial J'} \qquad (12)$$

Equation (7) can also be linearized in $h$ at $J = J_C$. Using the fact that $\zeta_m$ is an even function of $h$ we obtain

$$\left[\frac{m}{m'}\right]^{2Y_h} = \frac{\partial^2 \zeta_m/\partial h^2}{\partial^2 \zeta_{m'}/\partial h'^2} \qquad (13)$$

The Mathematica package is used to diagonalize the reduced transfer matrix [25], from which the eigenvalues are calculated with a high precision for different values of $J_x, J_y$ and $J_z$. Our results for $J_C$, $Y_t$ and $Y_h$ are shown in the following tables. All the critical exponents can be calculated by the following relations

$$Y_t = 2a_t \quad , \quad Y_h = 2a_h \qquad (14)$$

$$\alpha = 2 - \frac{1}{a_t} \quad , \quad \beta = \frac{1-a_h}{a_t} \qquad (15)$$

$$\gamma = \frac{2a_h - 1}{a_t} \quad , \quad \delta = \frac{a_h}{1-a_h} \qquad (16)$$

Tables(1,2,3,4): Calculated critical properties of the anisotropic two-layer Ising model.

Table (1)

| $m/m'$ | $J_y : J_x : J_z$ | $J_C$ | $Y_t$ | $Y_h$ |
|---|---|---|---|---|
| 4/6 | | 0.360525 | 1.006837 | 1.880667 |
| 4/8 | | 0.348146 | 1.004480 | 1.878220 |
| 4/10 | | 0.340260 | 1.003337 | 1.876550 |
| 6/8 | 1:1:1 | 0.335967 | 1.002750 | 1.875640 |
| 6/10 | | 0.330325 | 1.001500 | 1.874148 |
| 8/10 | | 0.324720 | 1.000142 | 1.872350 |
| Extrapolated | | 0.316046 | 1.001592 | 1.875814 |
| Exact 2 - D Ising Model | | | 1.000000 | 1.875000 |

Table (2)

| $m/m'$ | $J_y : J_x : J_z$ | $J_C$ | $Y_t$ | $Y_h$ |
|---|---|---|---|---|
| 4/6 | | 0.240350 | 1.006837 | 1.880202 |
| 4/8 | | 0.232097 | 1.004814 | 1.878219 |
| 4/10 | | 0.226840 | 1.003340 | 1.876554 |
| 6/8 | 1:1.5:1.5 | 0.223978 | 1.002751 | 1.875660 |
| 6/10 | | 0.220216 | 1.001490 | 1.874134 |
| 8/10 | | 0.216480 | 1.000146 | 1.872251 |
| Extrapolated | | 0.210698 | 1.001586 | 1.875808 |
| Exact 2 - D Ising Model | | | 1.000000 | 1.875000 |

Table (3)

| $m/m'$ | $J_y : J_x : J_z$ | $J_C$ | $Y_t$ | $Y_h$ |
|---|---|---|---|---|
| 4/6 | | 0.721055 | 1.006837 | 1.880208 |
| 4/8 | | 0.696291 | 1.004813 | 1.878219 |
| 4/10 | | 0.680520 | 1.003335 | 1.876555 |
| 6/8 | 1:0.5:0.5 | 0.671934 | 1.002750 | 1.875640 |
| 6/10 | | 0.660650 | 1.001500 | 1.874147 |
| 8/10 | | 0.649441 | 1.000146 | 1.872462 |
| Extrapolated | | 0.632094 | 1.001592 | 1.875818 |
| Exact 2 - D Ising Model | | | 1.000000 | 1.875000 |

Table (4)

| m/m′ | $J_y : J_x : J_z$ | $J_C$ | $Y_t$ | $Y_h$ |
|---|---|---|---|---|
| 4/6 | | 0.180262 | 1.006837 | 1.880667 |
| 4/8 | | 0.174072 | 1.004820 | 1.878202 |
| 4/10 | | 0.170130 | 1.003342 | 1.876554 |
| 6/8 | 1 : 2 : 2 | 0.167983 | 1.002751 | 1.875627 |
| 6/10 | | 0.165162 | 1.001494 | 1.874133 |
| 8/10 | | 0.162360 | 1.000163 | 1.872252 |
| Extrapolated | | 0.158024 | 1.001590 | 1.875805 |
| Exact 2-D Ising Model | | | 1.000000 | 1.875000 |

Table (5): All of the critical exponents for the two-layer Ising model.

| $J_y : J_x : J_z$ | α | β | γ | δ |
|---|---|---|---|---|
| 1:1:1 | 0.003179 | 0.12399 | 1.7488 | 15.1049 |
| 1:1.5:1.5 | 0.003167 | 0.12399 | 1.7488 | 15.1041 |
| 1:0.5:0.5 | 0.003179 | 0.12398 | 1.7488 | 15.1054 |
| 1:2:2 | 0.003175 | 0.12399 | 1.7488 | 15.1037 |
| Exact 2-D Ising Model | 0 | 0.125 | 1.75 | 15 |

The calculation have been done for a lot of points along the critical line for the symmetric two layer Ising model with $J_x$, $J_y$ for the coupling constants in the $x$ and $y$ directions in each layer (parallel couplings in the up and down layers are equal) and with inter layer coupling constant $J_z$. For extrapolating the data a simple method can be used, for example in table(3),

$$\frac{J_x}{J_y} = 0.5 \qquad \frac{J_x}{J_z} = 1 \qquad \frac{J_y}{J_z} = 2 \qquad (17)$$

The following code in Mathemathica can be used to obtain $Y_h$.

$$Fit[\{\{\frac{4}{10}, 1.876555 \ Log[\frac{4}{10}]\}, \{\frac{6}{8}, 1.87564 \ Log[\frac{6}{8}]\}, \{\frac{6}{10}, 1.84147 \ Log[\frac{6}{10}]\},$$
$$\{\frac{8}{10}, 1.872462 \ Log[\frac{8}{10}]\}\}, \{Log[x]\}, x] \qquad (18)$$

which yields $Y_h = 1.87582$.

III. **Conclusion**

It is clear that the two layer Ising model is in the same universality class as the two dimensional Ising model, the error is less than 0.15%. We conclude that the critical exponents are constant along the critical line and the model is in the same universality class as the two dimensional Ising model.

**Acknowledgements**

Our thanks go to the Isfahan University of Technology and Institute for Studies in Theoretical Physics and Mathematics for their financial support.